# Glucose Levels Sensing using Whispering Gallery Modes at mm-Wave Band

Ala Eldin Omer, George Shaker, and Safieddin Safavi-Naeini
Centre for Intelligent Antenna and Radio Systems (CIARS), University of Waterloo, Waterloo, Canada
Email: aeomomer@uwaterloo.ca

*Abstract*— In this study, an integrated low-cost and complexity mm-wave structure of Whispering Gallery Mode (WGM) is proposed for sensing the glucose levels in mimicking aquatic solutions of concentrations similar to type 2 diabetics. The bio-sensor is composed of a curved dielectric waveguide coupled to a dielectric disc resonator that is loaded with the glucose sample under test. The intense WGM field induced towards the boundary of the DDR is exploited to detect the slight variations in the dielectric properties of different glucose levels via tracing the variations in the magnitude of the transmission coefficient $S_{21}$ in the mm-wave frequency band (49 – 70 GHz). The WGM resonator shows a high sensitivity performance (0.025-0.077 dB/(mg/dL)) at the lower-order modes $WGH_{600}$ and $WGH_{700}$ as demonstrated by simulations in a 3D full-wave EM solver (Ansys HFSS).

## I. Introduction

Millimeter-wave (mm-wave) frequencies have shown to be promising in developing various biomedical appliances due the enhanced security, available spectrum, reduced interference, and miniaturized sizing for fabricated devices [1]. The higher spatial resolution than microwaves, better sensitivity to the lossy water content, and less sensitivity to water contaminants; are all features that make the mm-wave heavily investigated by researchers to promote portable devices for sensing glucose concentrations [2]. This is becoming particularly important because diabetics need to continually monitor their blood glucose levels in order to meet certain glycemic targets and control it accordingly through remedies to avoid the susceptibility of any serious health complications [3][4]. Generally, typical glucose levels in human beings blood for type 2 diabetes widespread among 90% of people normally vary in a small range of milligrams per millilitre, specifically from 0.72 to 1.4 mg/ml before having a meal and should be less than 180 mg/dL within 1 – 2 hours after a meal [5].

The finger-pricking glucometer device has been the primary technique used by diabetics to check their blood glucose levels despite the many limitations of being painful, invasive, and costly to use [6]. To overcome these limitations, researchers have conducted thorough investigations over the last decade on alternative, non-invasive glucose detection (NGD) methods, including the optical techniques that can identify glucose concentrations through optical parameters, such as: optical coherence tomography, near-infrared and Raman spectroscopy, light scattering, fluorescence measurements, and more [7–9]. Absorbance spectroscopy techniques, such as red/near infrared [10] and mid infrared [11] spectroscopy, that use the scattering of light on biological tissue to detect optical signatures of glucose in blood. The radio frequency (RF) and microwave techniques have also been widely researched in [12–20] that show a promising correlation between the measured electromagnetic (EM) properties of blood and its glucose content. For instance, the dielectric properties of blood glucose in superficial vessels were measured using split ring resonators [21-22], microstrip patch antennas [23-24], vector network analyzers (VNA) [25], and waveguides [26-27].

Featured dielectric waveguides emerge as a reasonable substitution to overcome the limitations of the traditional metallic waveguides in manufactural inaccuracies and conductive losses at mm-wave frequencies and beyond. They offer low-cost, low-loss and high confinement wave transmission inside whispering gallery mode (WGM) resonators that demonstrate high sensing capabilities [28]. WGMs are high-order modes characterized by the unique field distribution inside a dielectric disk/ring resonator (DDR/DRR) where most of the energy is confined towards the resonator boundaries. Compared to the conventional dielectric resonators of lower-order modes [29], WGM resonators have higher Q-factor resonances with a relatively large size (in terms of wavelength) that can tolerate the dimensional errors in their fabrication processes. For all WGM modes with small axial propagation constant, the coupled waves in WGM disc/ring resonators travel azimuthally close to the resonator circumference boundary. The type of WGM is primarily determined by the physical dimension of the resonator and its method of excitation. Among the many WGMs, $WGE_{n00}$ and $WGH_{n00}$ modes are more desirable for effective sensing where the EM energy is confined and concentrated in small regions near the frontier of the DDR. Subsequently, the resonance characteristics of the DDR become very sensitive to any small perturbation close to its boundary region making the resonator capable of differentiating various materials with very small differences in EM properties.

In this paper, we propose a reliable low-cost mm-wave integrated WGM-based glucose bio-sensor that consists of a dielectric disc resonator (DDR) coupled to a curved dielectric image waveguide (DIG). The proposed sensor aims at monitoring the variations in blood glucose levels on a non-invasive continuous manner in the mm-wave regime. This is illustrated via the interaction between the lossy glucose sample loaded on top of the DDR and the WGM field that is strongly coupled towards the boundaries of the loaded sample, and hence reflected into traceable variations in the $S_{21}$ magnitude for tiny variances in dielectric properties of different glucose concentrations.

This work was supported by the CREATE grant via the Natural Sciences and Engineering Research Council of Canada (NSERC) and by Schlegel UW Research Institute of Aging (UW-RIA).

## II. DIELECTRIC PROPERTIES OF GLUCOSE SOLUTIONS

For preliminary investigations on non-invasive glucose detection using RF sensors, aqueous glucose solutions can be utilized in the RF experiments to mimic blood samples at disparate glucose concentrations while assuring sufficient repeatability of the collected measurements. This approximation is valid since water contributes high percentage ($\approx$ 50%) of the entirety volume of human blood that contains other vital constituent components at different proportions (e.g. glucose, Na, Ca, K, Cl, etc) [30]. Aqueous solutions introduce dispersion to the EM waves propagated through their media at different frequencies. The dielectric response in the frequency domain of tissues having high water content can be characterized by the Debye Relaxation Model [31].

A few research studies have focused on extracting the permittivity measurements for liquid samples of relatively high glucose concentrations at higher frequency ranges. In [32] a waveguide transmission probe was clamped onto the ear of a live anesthetic rat in order to observe the EM properties continuously on VNA. The authors noted a correlation between blood glucose concentrations and the absorption/reflection properties of the mm-wave transmitted in the Ka band (27 – 40 GHz) as demonstrated by the considerable changes in the absorption/reflection properties during the 0.5 mL intraperitoneal (IP) injection of glucose and insulin into the rat under experiment. The authors of [19] use different sensing techniques that employ both reflection and transmission modes to collect the permittivity measurements for aqueous glucose solutions in a frequency range up to 40 GHz for higher glucose concentrations from 0.15 wt% up to 0.5 wt%. The study in [12] considered the 50 – 75 GHz band for measurements of reflection coefficients using saline solutions of high concentrations (0.5 wt% and 3 wt%). A dependency between these concentrations and reflected coefficients is shown but not enough correlation is demonstrated even for the high glucose concentrations. Another study [33] in the frequency range (28 – 93 GHz) used the mm-wave spectroscopy in reflection mode to detect sugar in watery solutions, blood imitators (physiological solution: 0.9% NaCl in water), and blood while checking the sensitivity of their dielectric properties to glucose content at some concentrations ranging from 0.5 to 5% wt. The complex permittivity of the lossy medium under test is calculated from a simple structure of a plane-parallel matching plate made of a low-loss dielectric. The authors in [5] were able to demonstrate a clear linear correlation between transmitted EM energy in the mm-wave band (50 – 75 GHz) and the glucose concentrations in watery and salty solutions. Two open-ended V-band waveguides (WR-15) connected to a PNA were used to measure the dielectric properties, reflection ($S_{11}$) and transmission ($S_{21}$) coefficients. The tested samples have different types of glucose concentrations in a clinically relevant range between 0.025 wt% and 5 wt%. Transmission data achieved better results compared to reflection one especially in the 59–64 GHz and 69–73 GHz frequency ranges. A recent study in [24] proposed a mm-wave sensing system that consists of a pair of 60 GHz microstrip patch antennas placed across the interrogated blood sample. Measurements of the transmission coefficients are shown to be highly affected by changes in the permittivity properties along the signal path due to variations in glucose concentrations.

Following up the aforementioned efforts, we measure the dielectric properties (dielectric constant and loss tangent) of assorted glucose-water solutions prepared at similar concentrations of diabetes (0.7 – 1.2 mg/ml, in steps of 0.1 mg/ml) over the mm-wave spectrum from 50 to 67 GHz using a commercial coaxial probe kit (DAK-TL) connected to a vector network analyzer (VNA) to identify the amount of change in dielectric constant and loss tangent, due different concentrations of interest [34]. The collected measurements plotted in Fig. 1 show that our concentrations of interest are better distinguishable in this high frequency range where the dielectric constant is decreasing exponentially as the frequency increases and the loss tangent is increasing nearly in a linear pattern up to 60 GHz then gradually drops. Most notably, with an increased level of glucose content in water, we find the dielectric constant slightly increases while the loss tangent is tenuously decreasing like the pattern in [5].

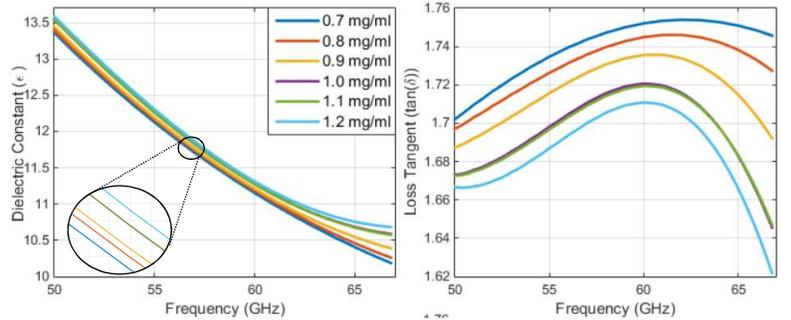

Fig. 1. Dielectric measurements of glucose solutions of different concentrations.

Measured dielectric parameters are fitted to a single-pole Debye model of independent coefficients as listed in Table I, then used to analyze the sensitivity performance of the designed WGM sensor when loaded with different glucose concentrations via integrating their dielectric values at a certain frequency.

Table I. Extracted Debye coefficients for each glucose concentration

| Glucose samples mg/ml | $\epsilon'_\infty$ | $\epsilon_s$ | $\tau$ (ps) |
|---|---|---|---|
| 0.7 | 5.67 | 80.65 | 9.42 |
| 0.8 | 5.76 | 81.028 | 9.49 |
| 0.9 | 5.92 | 81.95 | 9.65 |
| 1.0 | 6.13 | 82.95 | 9.82 |
| 1.1 | 6.147 | 83.208 | 9.848 |
| 1.2 | 6.267 | 83.79 | 9.948 |

## III. THE PROPOSED MM-WAVE SENSOR

The WGM sensor composes of a curved alumina image guide ($\epsilon'_r = 9.2, tan\delta = 0.0005$) coupled to a disc resonator made of a low-loss Silicon material ($\epsilon'_r = 11.6, tan\delta = 0.001$). The DDR has a radius $R = 3.42$ mm and height $h = 0.5$ mm. The DIG is designed with a curvature ($R = 4$mm, $\theta = 45°$) in the area around the DDR to realize a packaged system. It has a cross section of 1x0.8 mm$^2$ and total length of 50 mm with linear tapers at both ends to reduce the insertion losses. The dielectric guide is excited by the standard rectangular metallic waveguide (WR-15) to support the dominant mode ($E_z^{11}$) operation in the mm-wave band 49–70 GHz, resulted into operating the DDR in WGH mode. The proposed design of the sensor structure shown in Fig. 2 is simulated and optimized with the parameters shown in the Table II using a 3D EM solver. A plexiglass container ($\epsilon'_r = 3.4, tan\delta = 0.001$) of height $h_c$, side and bottom thickness $t_1$ and $t_2$ is used to hold the glucose samples on top of the DDR. Five distinct WGH resonance modes are observed in the entire mm-wave band as shown in Fig 3 with small black circles. The corresponding $S_{21}$ curves are plotted when the DDR is unloaded, loaded with empty container, and loaded with a 0.7 mg/ml glucose sample. Figure 3 also depicts shift in the frequencies of the WGH resonances when the DDR is loaded with the empty container and another shift when the glucose sample is loaded. However, varying the glucose levels would only change $|S_{21}|$.

The parametric sweep function in HFSS is used to optimize the sensitivity and selectivity performance of the integrated sensor when loaded with glucose samples of parameters in Table I inside the plexiglass container. We aim at creating a critical coupling between the loaded resonator and the waveguide at the different supported WGH modes to realize a high Q-factor for better sensitivity. Therefore, for a given WGH mode we sweep the gap parameter g from 0.1mm to 0.8mm in steps of 0.01mm when the structure is loaded with different glucose samples at the corresponding resonance frequency. The results have shown that the resonator will have the same critical coupling gap when loaded with the different glucose samples. However, this critical coupling condition is varying from one WGH mode to another as shown in Table III. Another parameter that affects the coupling between the DDR and glucose sample is the bottom thickness of the container $t_2$. The container is designed with $t_2$ = 1.5 mm that resulted into better coupling with the glucose sample when $t_2$ is swept from 0 mm to 3mm in steps of 0.25 mm. The side wall thickness has no significant effect and is set to 0.1mm.

Finally, we study the sensitivity behavior of the DDR under the critical coupling condition for each supported WGH mode when loaded with different glucose concentrations 0.7, 0.8, 0.9, 1.0, and 1.1. Figure 4 shows the $S_{21}$ plots of the different glucose samples at $WGH_{600}$ and $WGH_{700}$. It is noticed that the first lower modes $WGH_{600}$ and $WGH_{700}$ achieve higher sensitivity for small changes in glucose levels compared to the higher order modes $WGH_{800}$, $WGH_{900}$ in the mm-wave range. For instance, when the sensor is operating at $WGH_{600}$ and $WGH_{700}$ the $S_{21}$ varies by 0.76 dB and 0.25 dB, respectively, when the glucose concentration increases by 0.1 mg/ml. However, for a similar glucose change we find the $S_{21}$ varies by less than 0.04 dB for $WGH_{800}$ mode. This tendency in sensitivity can be interpreted by the nature of the interaction between the glucose sample and the coupled WGM field that, for higher-order-modes, have the particularity of being very confined around the DDR boundary with a very tenuous evanescent field tail which is not enough to sense the small external perturbations of varying the glucose levels outside the DDR. The sensitivity $\Delta S_{21}/\Delta C$ in dB/(mg/dL) is computed in Table III for three different WGH mode at 49.9, 54.5, and 58.9 GHz, respectively. This high level of sensitivity is superior to recent works in the literature [35-37].

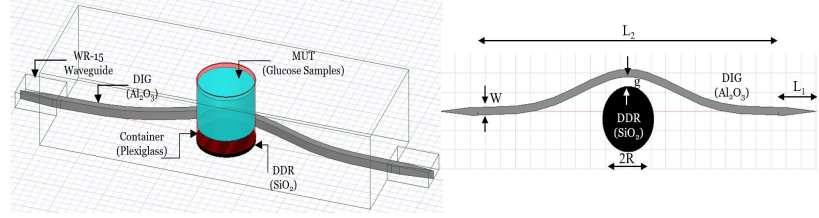

Fig. 2. The proposed sensor structure with the design parameters, 3D view (left) and top view (right).

Table II. Design parameters of the proposed sensor structure shown in Fig. 2

| Parameter (mm) | W | H | $L_1$ | $L_2$ | R | h | $h_c$ | $t_1$ | $t_2$ | $h_{MUT}$ |
|---|---|---|---|---|---|---|---|---|---|---|
| | 0.8 | 1.0 | 5 | 40 | 3.42 | 0.5 | 7 | 0.1 | 1.5 | 5 |

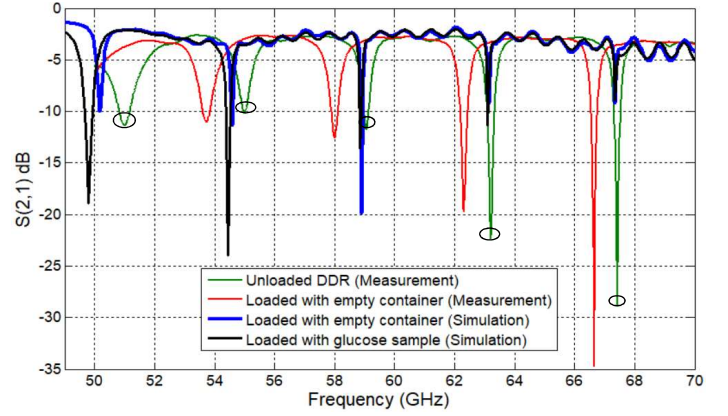

Fig. 3. Measurement and simulation results of $S_{21}$ for different WGH modes.

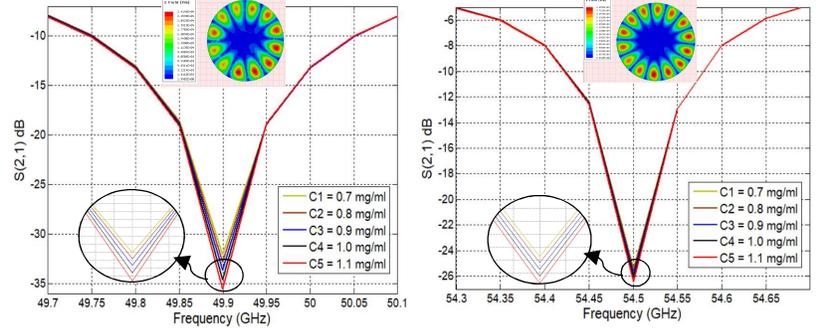

Fig. 4. Simulation results of $S_{21}$ responses and electric-field distributions at the bottom of the DDR at resonances of $WGH_{600}$ (left) and $WGH_{700}$ (right).

Table III. Critical coupling gap and sensitivity for each WGM mode in the mm-wave band

| WGM Mode | Critical Coupling Gap g (mm) | Sensitivity (dB/(mg/dL)) |
|---|---|---|
| $WGH_{600}$ | 0.42 | 0.077 |
| $WGH_{700}$ | 0.45 | 0.025 |
| $WGH_{800}$ | 0.45 | 0.0104 |

### CONCLUSION

A reliable non-invasive mm-wave integrated glucose bio-sensor is proposed in this paper. The sensor exploits the WGM fields excited on the boundaries of a DDR via an Alumina-based curved image guide to detect the small variations in the dielectric properties of the loaded glucose sample. The sensor achieves a high level of sensitivity when the lower order modes $WGH_{600}$ and $WGH_{700}$ are excited inside the resonator as demonstrated by HFSS simulations in the mm-wave band.